\documentclass[11pt]{article}
\usepackage{graphicx}

\newcommand{\BABARPubYear}    {01}

\newcommand{\BABARConfNumber} {08}
\newcommand{\SLACPubNumber} {8979}
\newcommand{\LANLNumber} {0109034}

\newcommand{\petap}{p_{\eta'}^{*}}
\input babarsym

\long\def\inst#1{\par\nobreak\kern 4pt\nobreak
    {\it #1}\par\vskip 10pt plus 3pt minus 3pt}

\begin{document}
\pagestyle{empty}
\begin{flushright}
\babar-CONF-\BABARPubYear/\BABARConfNumber \\
SLAC-PUB-\SLACPubNumber \\
hep-ex/\LANLNumber \\
September, 2001 \\
\end{flushright}

\par\vskip 5cm

\begin{center}
\Large \bf Study of semi-inclusive production of $\eta'$ mesons in $B$ decays
\end{center}
\bigskip

\begin{center}
\large The \babar\ Collaboration\\
\mbox{ }\\
\today
\end{center}
\bigskip \bigskip

\begin{center}
\large \bf Abstract
\end{center}
We report a measurement of the rate for $B \to \eta' X_s$ transitions where
the $\eta'$ meson has center-of-mass momentum in the range
2.0 to 2.7\gevc\ and $X_s$ represents a system comprising 
a kaon and up to four pions. Our study is based on 
22.2 million $B\overline{B}$ pairs collected at the $\Upsilon(4S)$ 
with the \babar\ detector at the Stanford Linear Accelerator Center.
We find $\mathcal{B}(B \to \eta'
X_s)=(6.8^{+0.7}_{-1.7}(stat)\pm1.0(syst)^{+0.0}_{-0.5}(bkg))\times10^{-4}$
assuming that the signal is due to $b \to sg^*$ transitions.

\vfill
\begin{center}
Submitted to the 
9$^{th}$ International Symposium on Heavy Flavor Physics \\
9/10---9/13/2001, Pasadena, CA, USA
\end{center}

\vspace{1.0cm}
\begin{center}
{\em Stanford Linear Accelerator Center, Stanford University, 
Stanford, CA 94309} \\ \vspace{0.1cm}\hrule\vspace{0.1cm}
Work supported in part by Department of Energy contract DE-AC03-76SF00515.
\end{center}

\newpage
\pagestyle{plain}

\begin{center}
\small

The \babar\ Collaboration,
\bigskip

B.~Aubert,
D.~Boutigny,
J.-M.~Gaillard,
A.~Hicheur,
Y.~Karyotakis,
J.~P.~Lees,
P.~Robbe,
V.~Tisserand
\inst{Laboratoire de Physique des Particules, F-74941 Annecy-le-Vieux, France }
A.~Palano,
A.~Pompili
\inst{Universit\`a di Bari, Dipartimento di Fisica and INFN, I-70126 Bari, Italy }
% ...

\end{center}\newpage

\section{Introduction}
\label{sec:Introduction}

The study of $B$ decay modes involving 
gluonic penguin transitions $b \to sg^*$ is important  
both for obtaining a better understanding of the mechanisms contributing to $B$ decays
and as a sensitive place to search for direct \CP violation
effects. Exclusive modes such as $B^+ \to
\eta' K^+$ or $B^0 \to K^+ \pi^-$  are expected to be
dominated by penguin amplitudes \cite{ref:Neubert} and may show
direct \CP-violating charge asymmetries \cite{ref:Ali}. 
The study of a collection of decay modes
$B \to \eta' X_s$, where $X_s$ denotes a
set of decay particles containing an $s$ quark, is another
attractive method for obtaining inclusive information about $b \to sg^*$ transitions. 

\begin{figure}[!htb]
\begin{center}
\includegraphics[scale=0.5]{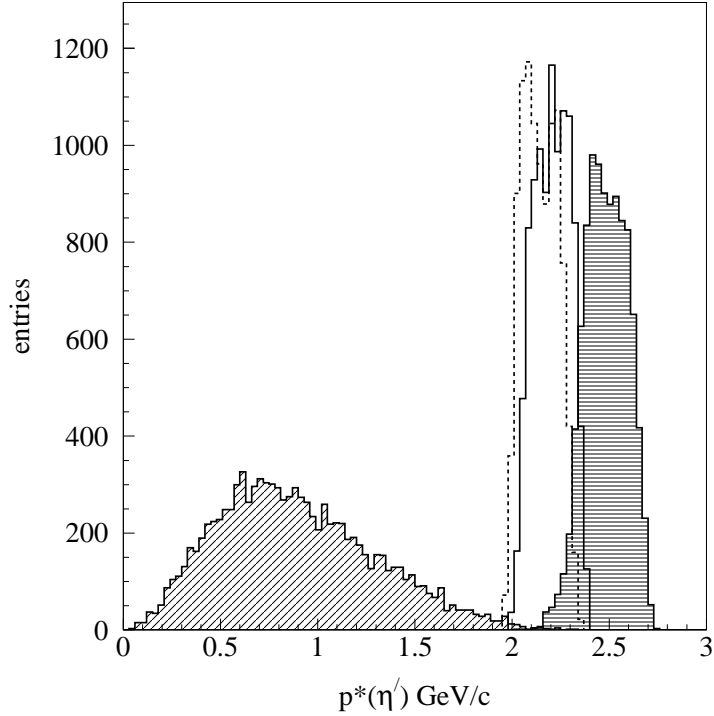}
\caption{Distribution of $\petap$ from Monte Carlo simulation of various
$b \to sq\overline{q}$ processes, such as $B \to \eta'K$,
$\eta'K^*$ (horizontally hatched histogram), $\eta'$ from a mixture of 
$B\overline{B}$ sources dominated by $b\to c$ cascade contributions
(diagonally hatched histogram), $\eta'D^0$ (open histogram)
and $\eta'D^{*0}$ (dashed histogram) samples.} \label{Fi:cmsp}
\end{center}
\end{figure}

In this paper, we report an application of this last approach 
to obtain a semi-inclusive measurement of the rate for $B\to\eta' X_s$ for
center-of-mass momentum $\petap$ of the $\eta'$ ranging from 2.0 to 2.7\gevc. In this
high momentum interval,
$\eta'$ production 
from $b \to c \to \eta'$ cascades, such as $B \to D_sX$ with $D_s \to \eta'X$, 
$B \to D^+X$ with $D^+ \to \eta'X$, $B \to D^0X$ with $D^0 \to \eta'X$, 
$B \to \Lambda_c X$ with $\Lambda_c \to \eta'X$, is suppressed,
although
it is important to note that other $B$ decay processes, such as
$b \to u$ decays ($B \to \eta' \pi$, $\eta' \rho$, $\eta' a_1$) 
and internal spectator $\b \to c$ decays ($\Bzb \to \eta' D^{0(*)}$) 
may still contribute. 
Figure~\ref{Fi:cmsp} shows the momentum distribution for the 
relevant processes.
The various contributions can be distinguished by kaon
identification and by examining the mass spectrum of the system
recoiling against the $\eta'$ ($M(X_s)$). This is possible if
the recoiling hadronic system is reconstructed with a semi-inclusive set
of possible decay modes. CLEO used such a semi-inclusive 
reconstruction technique to obtain the branching fraction
$\mathcal{B}(B \to \eta'
X_s)=(6.2\pm1.6(stat)\pm1.3(syst)^{+0.0}_{-1.5}(bkg))\times
10^{-4}$ for $2.0<\petap<2.7\ \gevc$ \cite{ref:Cleopap}. This rate
is large in comparison
with available predictions \cite{ref:Datta}.
Our analysis uses a similar technique, although with
improved kaon identification and different Monte Carlo models.

\section{The \babar\ detector and dataset}
\label{sec:babar}

The data used in this analysis were collected with the \babar\ detector \cite{ref:babar} at the \pep2\ storage ring \cite{ref:pep2} located at the Stanford Linear Accelerator Center.
The study presented here is based on an integrated luminosity of 20.2\invfb\
corresponding to a sample of 22.2 million $B\overline{B}$ pairs at the $\Upsilon(4S)$ 
resonance (on-resonance) and 2.6\invfb\ collected with center of mass energy 40\mev 
below this resonance (off-resonance).

The asymmetric beam configuration in the laboratory frame provides a boost to the $\Upsilon(4S)$ increasing the momentum range of the $B$-meson decay products up to 4.3\gevc. Charged particles are detected and their momenta are measured by a combination of a silicon vertex tracker consisting  of five double-sided layers and a 40-layer drift chamber, both operating in a 1.5-T solenoidal magnetic field. Photons and electrons are detected by a CsI(Tl) electromagnetic calorimeter, which provides excellent angular and energy resolution with high efficiency for energies above 20\mev.

Charged particle identification is provided by measurements of the average
energy loss $\dedx$ in the tracking devices and the
Cherenkov angle in the detector for internally reflected Cherenkov
light (DIRC). The Cherenkov angle determination provides $K$-$\pi$
separation of better than $4\sigma$ below 3\gevc and
$2.5\sigma$ for the highest momenta.

\section{Event selection}

In order to select $B\overline{B}$ events, we use the following requirements:
\begin{itemize}
\item{There must be at least four charged tracks per event, in order to suppress low multiplicity events such as $e^+e^- \to e^+e^-$, $\mu^+\mu^-$, $\tau^+\tau^-$.}
\item{$R_2$, the ratio of the second to the zeroth Fox-Wolfram moment \cite{ref:FoxWolf}, must be less than 0.5. The distribution of this  variable, which varies from 0 to 1, is peaked at low values for spherical $B\overline{B}$ events while it is broader and shifted toward intermediate values for $e^+e^- \to q\overline{q}$ events ($q=u,d,s,c$).}
\item{ The total energy of charged and neutral particles is required to be at least 5\gev and below 15\gev. Events having a significant missing energy due to neutrinos, such as $\tau^+\tau^-$ events where 
one $\tau$ decay to a higher multiplicity
hadronic mode, such as $\tau \to \pi \pi \pi \nu_{\tau}$, are highly suppressed by this cut.}
\end{itemize}
The efficiency of this selection is $(98\pm1)$\% for $B\overline{B}$ Monte Carlo events.

\section{Semi-inclusive analysis}
\label{introsemetaP}

We form combinations of a charged kaon or a \KS, an $\eta'$ and as
many as four pions of which at most one is a $\pi^0$. Sixteen
decay modes and their charge conjugates are considered:
\begin{center}
\begin{displaymath}
\begin{array}{l}
B^{\pm}\rightarrow\eta'K^{\pm}(+\pi^0(+\pi^+\pi^-))                     \\
B^0/\overline{B}^0\rightarrow\eta'\KS(+\pi^0(+\pi^+\pi^-))              \\
B^{\pm}\rightarrow\eta'\KS\pi^{\pm}(+\pi^0(+\pi^+\pi^-))                \\
B^0/\overline{B}^0\rightarrow\eta'K^{\pm}\pi^{\mp}(+\pi^0(+\pi^+\pi^-))
\end{array}\nonumber
\end{displaymath}
\end{center}

The $\eta'$ is reconstructed in the $\eta \pi^+
\pi^-$ channel, from $\eta \to \gamma \gamma$ candidates only. 
We require a minimum energy of 50\mev for photons from $\eta \to \gamma \gamma$. 
Photons are rejected if they are consistent with originating from a $\pi^0$ having an energy above 200\mev.
Candidates with invariant mass within $3\sigma$ of the $\eta$ mass are kinematically 
fitted to the nominal $\eta$ mass and then combined with two charged tracks to form an $\eta'$ candidate.

To identify the $s$ quark in the $X_s$ system, we select
events either with a track consistent with a charged kaon or with
a reconstructed \KS in the $\pi^+\pi^-$ channel. The charged kaon
selection has been optimized to reduce background from $B \to
\eta' \pi$, $\eta' \rho$, and $\eta' a_1$ decays without losing too
much efficiency for the signal. For the \KS, we consider the two-dimensional 
angle $\alpha$ between the momentum of the \KS
candidate and the flight
direction, which is peaked near zero for a true \KS. 
We require $\alpha<0.05\rad$. Selected \KS candidates with invariant mass
within $3\sigma$ of the \KS mass are kinematically fitted to the nominal \KS mass.

\subsection{$B$ candidate selection}
Candidates for $B\to\eta' X_s$ are reconstructed in the 16 modes listed above. They are required to be consistent with a $B$ decay based on the 
energy substituted mass \cite{ref:babar},
$$M_{\mathrm{ES}} = \sqrt{ (s/2 + \mathbf{p}_i \cdot \mathbf{p}_B)^2 / E_i^2 - 
   \mathbf{p}_B^{\,2} }$$ 
and the energy difference
$$\Delta E = (E_i E_B - \mathbf{p}_i \cdot \mathbf{p}_B - s/2)/\sqrt{s}$$ 
where $\sqrt{s}$ is the total $e^+e^-$ center-of-mass energy.
The initial-state four-momentum $(E_i,\mathbf{p}_i)$
derived from the beam kinematics and the four-momentum
$(E_B,\mathbf{p}_B)$ of the reconstructed $B$ candidate 
are all defined in the laboratory.
The calculation of $M_{\mathrm{ES}}$ only involves the
three-momenta of the decay products, and is therefore
independent of the masses assigned to them.
An additional variable, the cosine of the angle in the center of mass frame
between the thrust axis of the $B$ candidate and the thrust axis of the 
remainder of the event, $\cos\theta_T$, is used to remove continuum background.
The selection criteria applied are the following:
\begin{itemize}
\item{$M_{ES}>5.265$\gevcc and $|\Delta E|<0.1$\gev, {\em i.e.,} consistent with the nominal $B$ mass
and known production energy; and}
\item{$|\cos\theta_T|<0.8$ to reduces the large continuum background, which is concentrated near $\cos\theta_T=\pm1$
while the expected signal is uniformly distributed in this
variable.}
\end{itemize}
For each event, we select only a single candidate in a given decay mode. The selected candidate is 
the one with the smallest $\chi^2$ defined as 
$\chi^2 = (M_{\rm ES}-M_B(\rm PDG))^2/\sigma(M_B^2) + \Delta E^2/\sigma(\Delta E^2)$, where the widths $\sigma(M_B^2)$ and $\sigma(\Delta E^2)$ are obtained
from Monte Carlo simulation.

\subsection{Determination of the $X_s$ mass spectrum}
\begin{figure}[!htb]
\begin{center}
\includegraphics[scale=0.5]{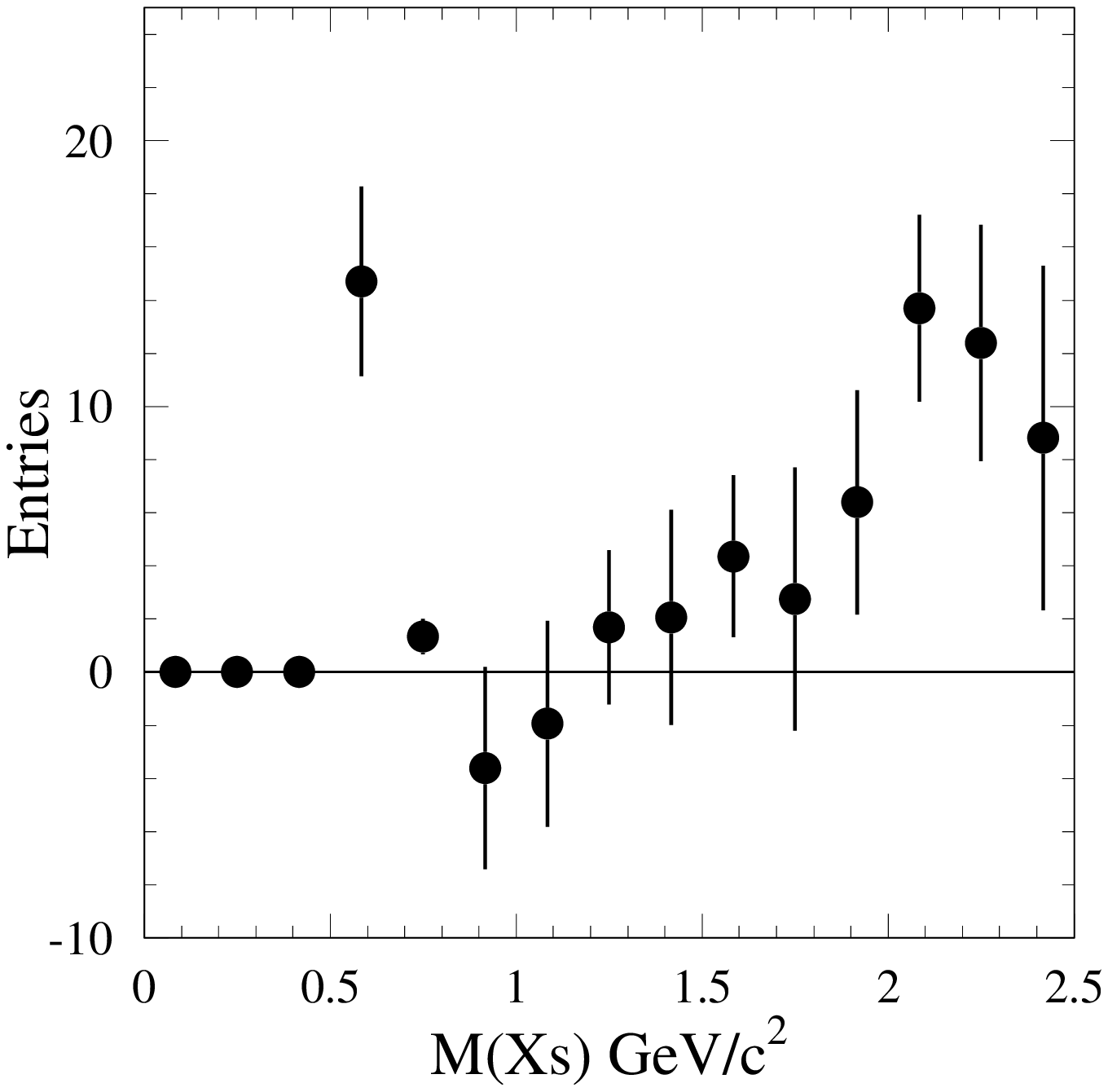}
\includegraphics[scale=0.5]{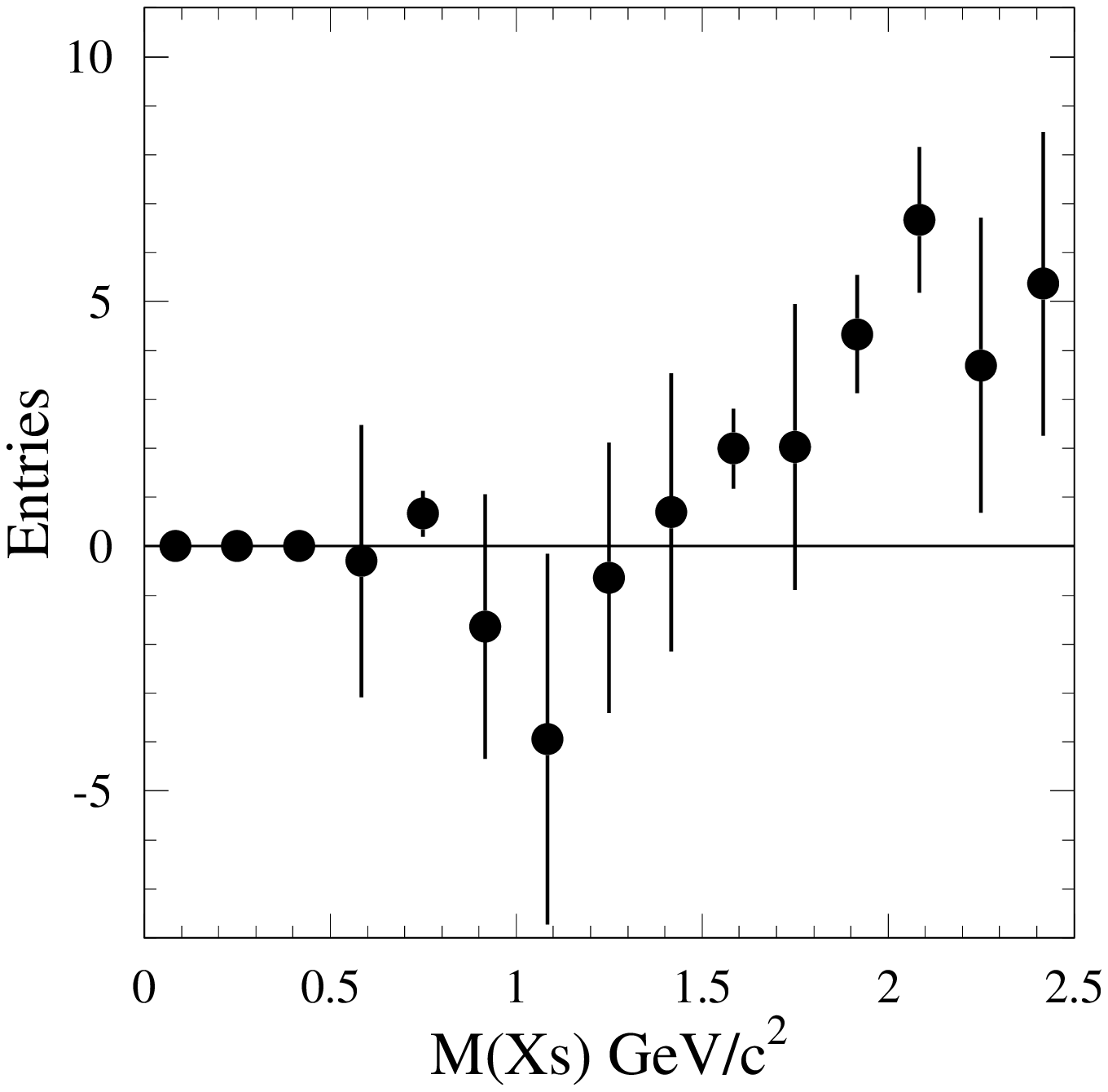}
\caption{Continuum-subtracted $M(X_s)$ spectrum for all $B$ modes (left)
and $B^0$ modes alone (right)} \label{Fi:mxsshape}
\end{center}
\end{figure}

\begin{figure}[!htb]
\begin{center}
\includegraphics[scale=0.55]{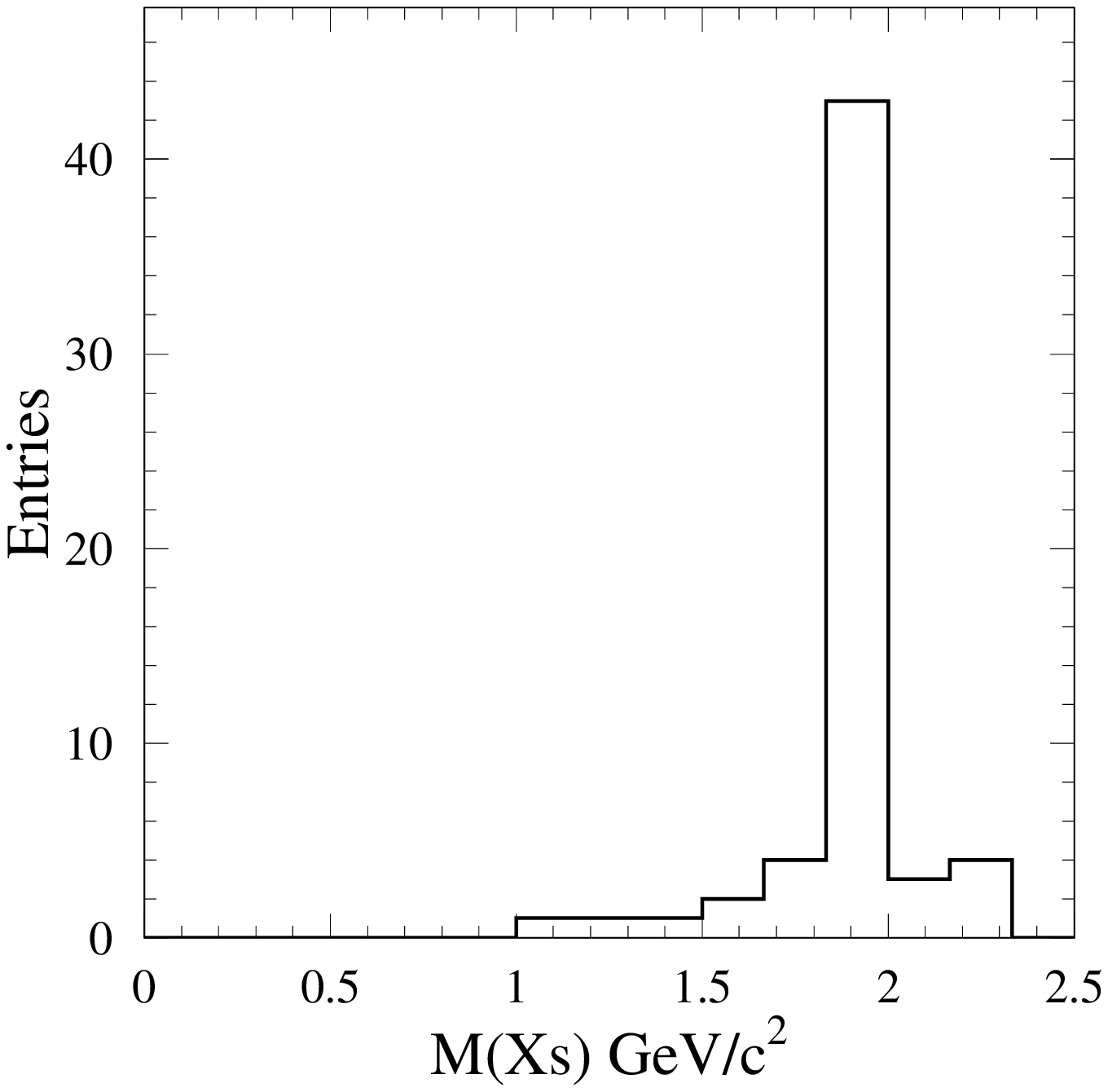}
\caption{$M(X_s)$ spectrum predicted from simulation of $\overline{B}^0\to\eta' D^0$ decays} \label{Fi:mxsetapd0}
\end{center}
\end{figure}

To explore the raw $X_s$ mass distribution, we first select the
$B$ candidates for which the mass of the
$\eta'$ daughter is within 3 sigma of the
known value, and
subtract the off-resonance contribution, 
rescaled by the luminosity ratio, from the
on-resonance distribution.
The resultant mass distributions for all $B$ modes and
separately for the \Bz\ modes are shown in Fig.~\ref{Fi:mxsshape}.
Both distributions can be seen to peak above 2\gevcc.
We can also examine the
$X_s$ mass spectrum for a possible signal for the internal
spectator charmed decays ($X_s = D^{0(*)}$ with $D^{*0} \to D^0
\pi^0$, $D^0 \to K \pi$, $K \pi\pi\pi$, $K\pi\pi^0$).
In particular, for the $B^0$ modes, there is no evidence for a narrow $D^0$ signal near 
1870\mevcc (see Fig.~\ref{Fi:mxsetapd0} for predicted distribution), although statistics are low.

\begin{figure}[!htb]
\begin{center}
\includegraphics[scale=0.55]{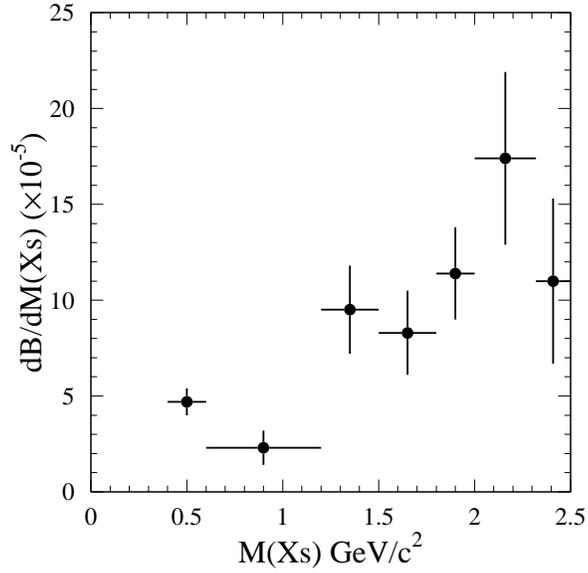}
\caption{Distribution of $d{\cal B}(B \to \eta' X_s)/dM(X_s)$
(statistical errors only) as a function of $M(X_s)$ for all $B$ modes.}
\label{Fi:mxslowbins}
\end{center}
\end{figure}

To obtain the decay $X_s$ spectrum we first fit
the $\eta'$ mass distribution in bins of $X_s$ mass.
For masses above 2.32\gevcc, 
corresponding to the kinematic limit $\petap<2\gevc$, 
the yield is dominated by the $b \to c \to \eta'$ contribution. 
The differential branching fraction for the region $M(X_s)<2.5$\gevcc, where we expect $b \to sg^*$ to be dominant, 
is shown in Fig.~\ref{Fi:mxslowbins}. The signal tends
to peak towards higher mass values, and remains
substantial between 2 and 2.5\gevcc. The experimental resolution 
for $M(X_s)$ has been estimated with the Monte Carlo simulation to be
80--90\mevcc around 1.5\gevcc, rising to 170--180\mevcc for 2\gevcc and above.


\begin{figure}[!htb]
\begin{center}
\includegraphics[scale=0.45]{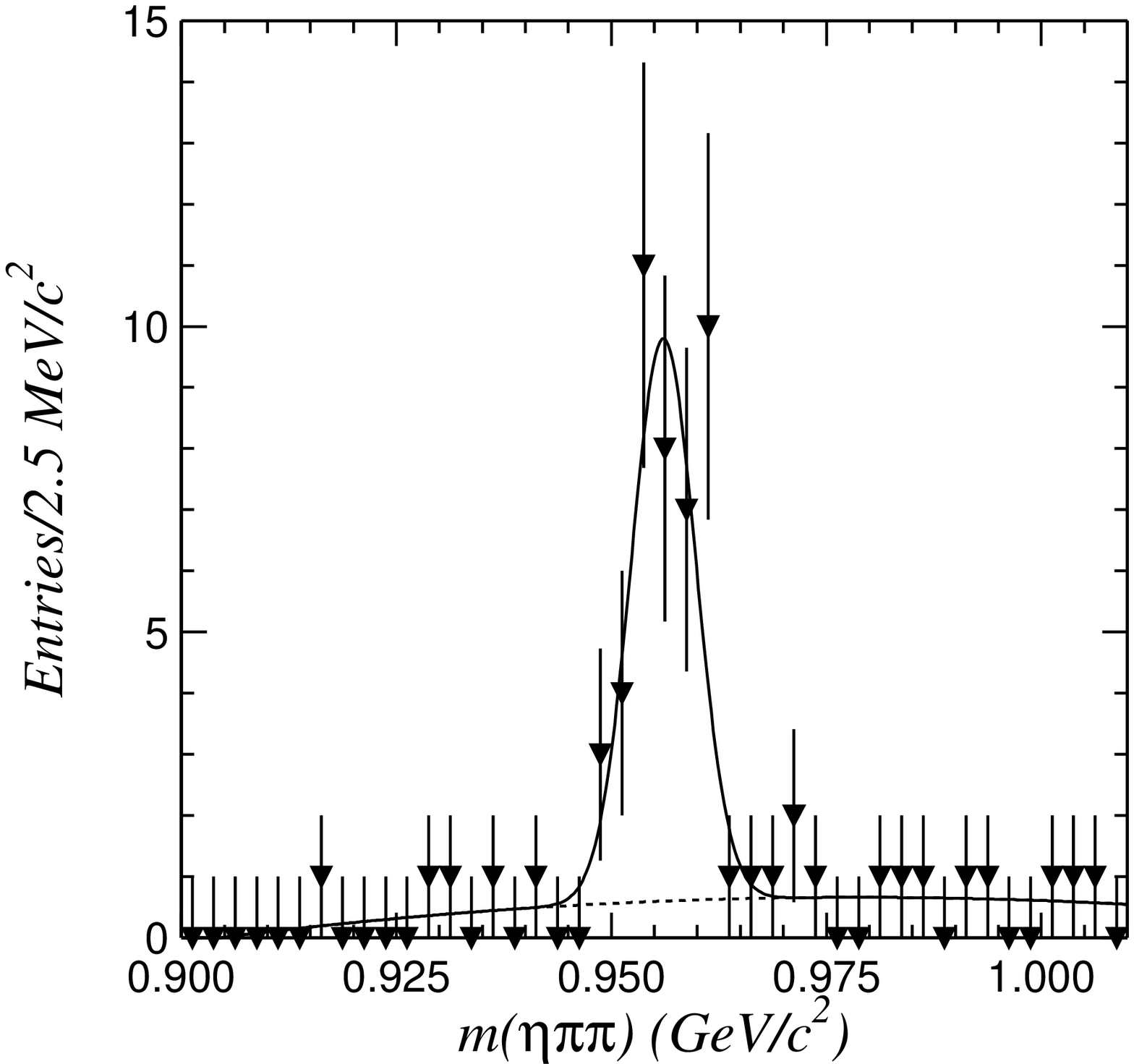}
\caption{$\eta'$ mass for $X_s=K^{\pm}$.}
\label{Fi:etapkch}
\end{center}
\end{figure}

Looking specifically at the two-body decay modes alone ($X_s=K^{\pm},\KS$), we find no significant
signal for $\eta'\KS$ but observe $36.2\pm6.6$ events
for $\eta' K^{\pm}$ (see Fig.~\ref{Fi:etapkch}). This
corresponds to a branching fraction of
$(5.6\pm1.0(stat))\times10^{-5}$, in good agreement with our
recent exclusive measurement of
$(7.0\pm0.8(stat)\pm0.5(syst))\times10^{-5}$ \cite{ref:oldetapk}, thereby
confirming the consistency of the semi-inclusive
method.

\subsection{Extraction of $\eta'$ signal}
\label{semetaP} 
\begin{figure}[!htb]
\begin{center}
\includegraphics[scale=0.4]{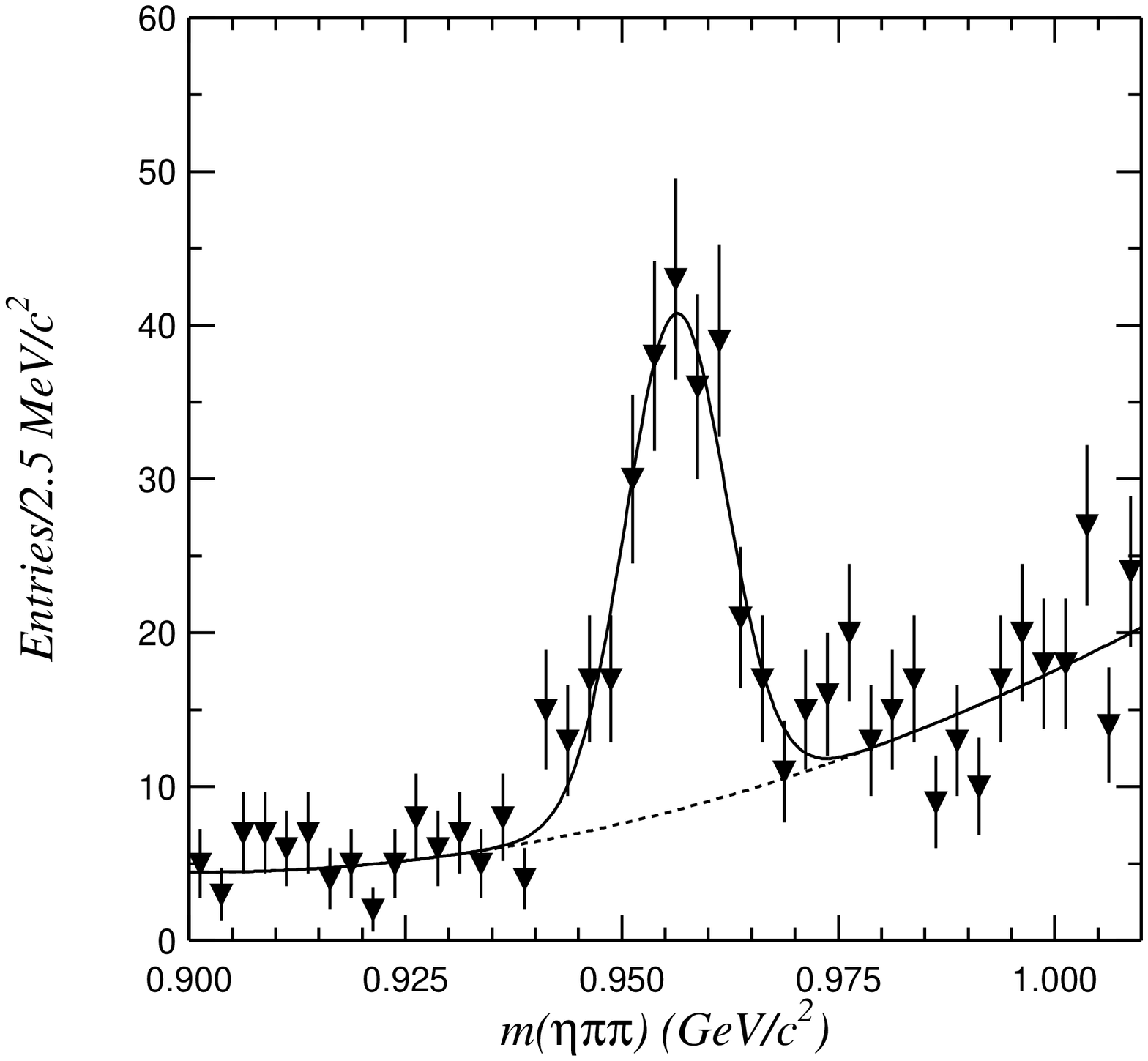}
\includegraphics[scale=0.4]{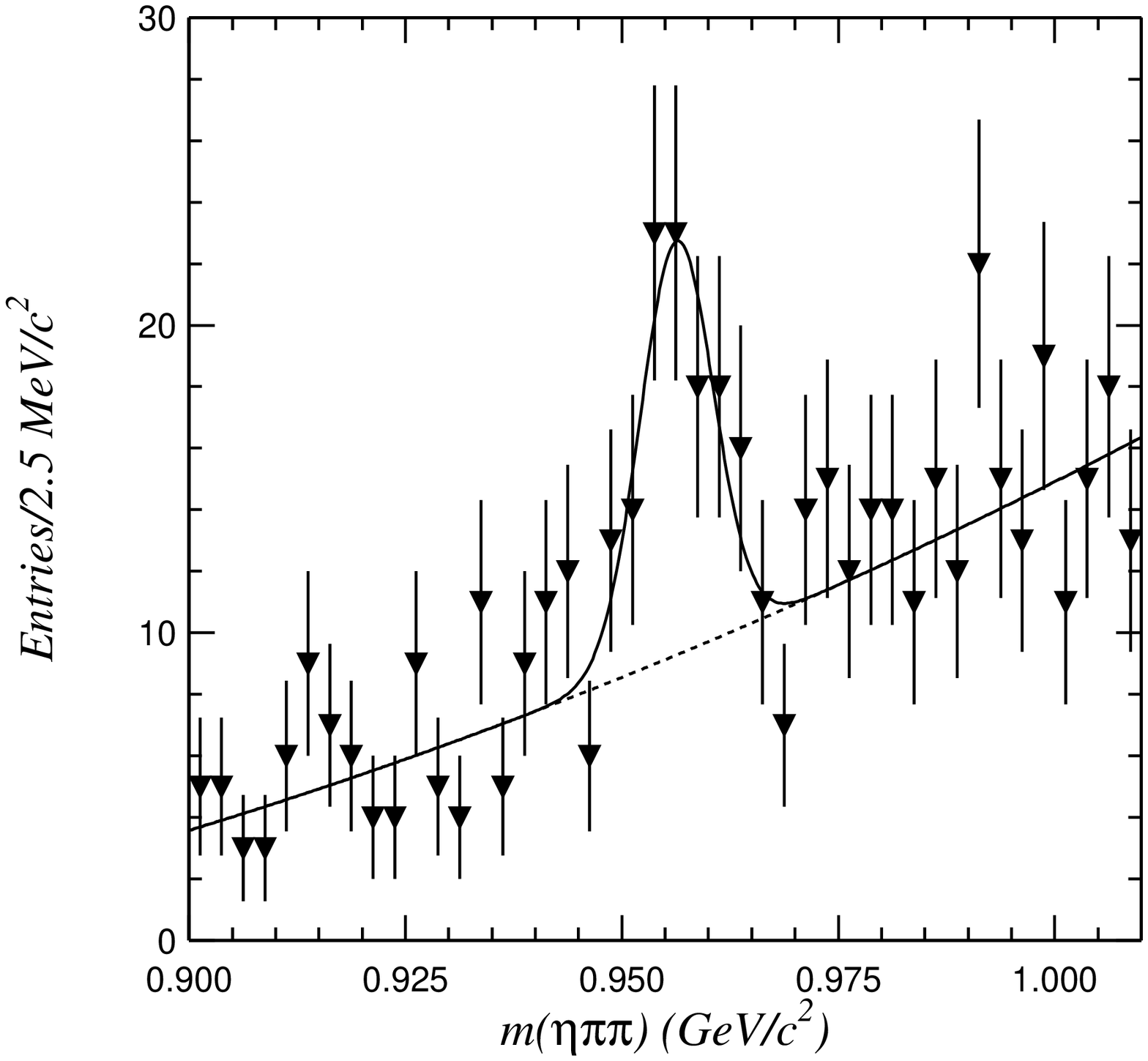}
\includegraphics[scale=0.53]{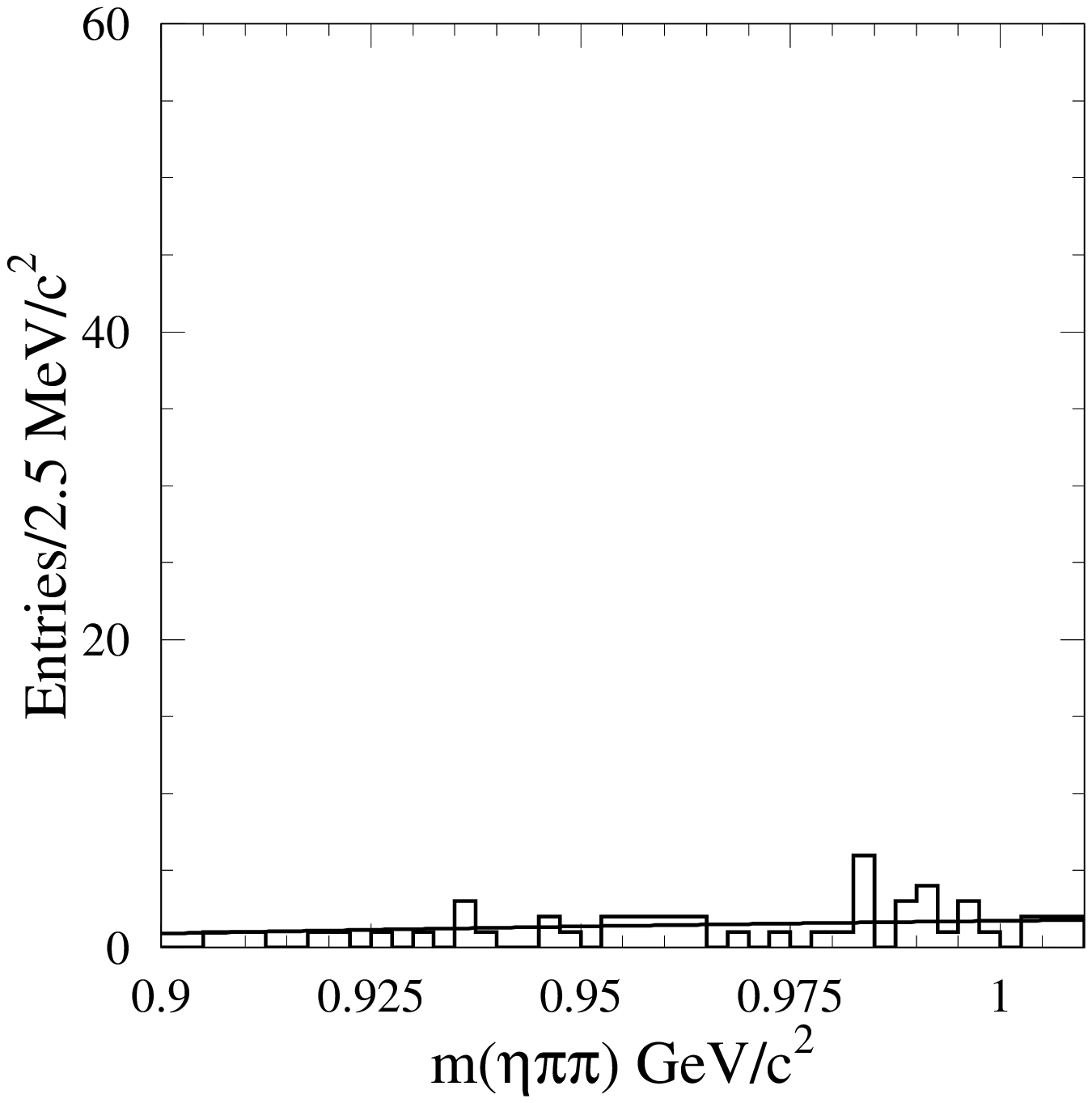}
\includegraphics[scale=0.53]{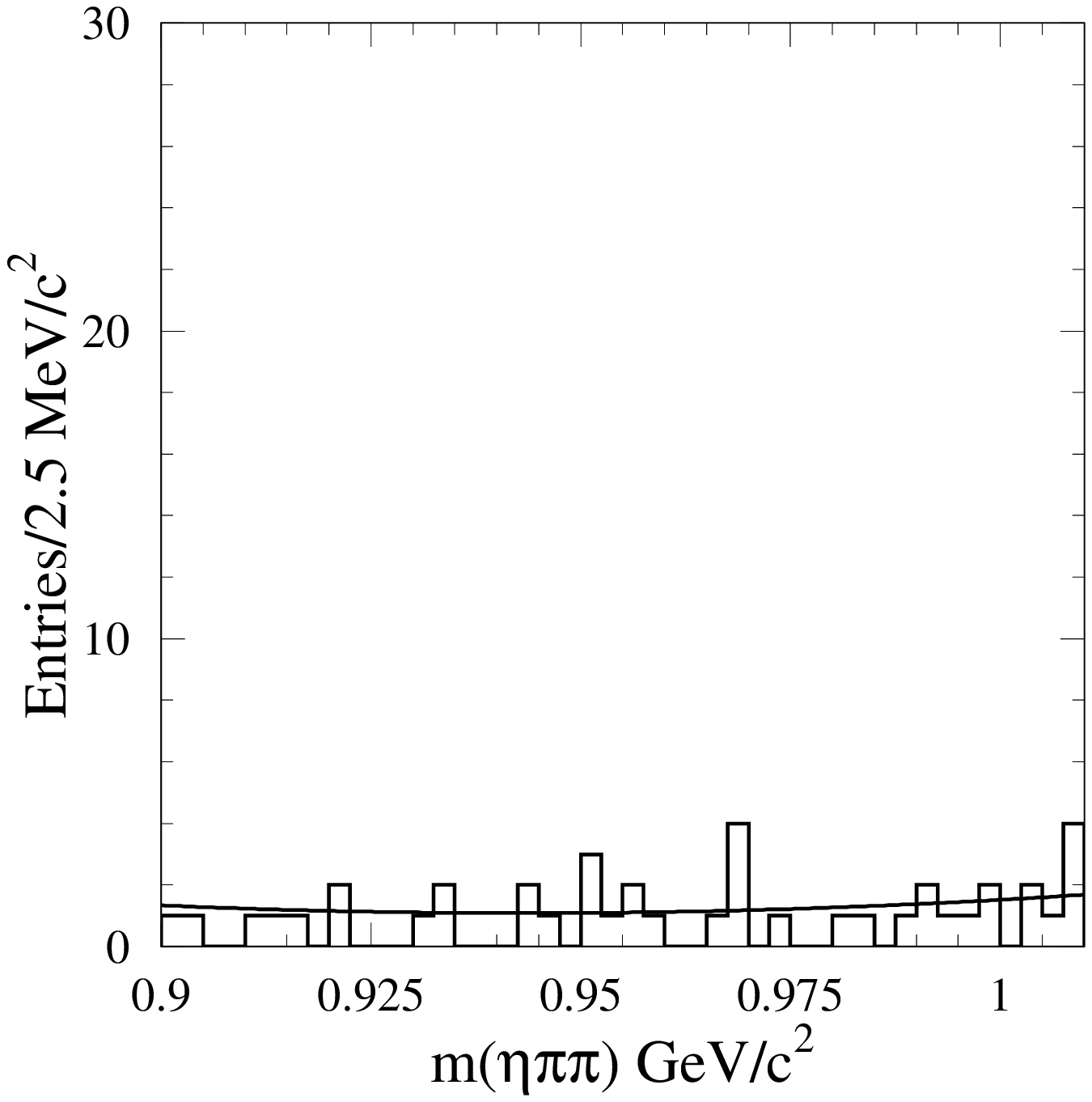}
\caption{Semi-inclusive $\eta'$ mass spectra for $K^{\pm}$ (upper left:
on-resonance, lower left: off-resonance) and \KS (upper right:
on-resonance, lower right: off-resonance) modes, 
in the momentum range $2.0<\petap<2.7\gevc $} \label{Fi:semExclsig}
\end{center}
\end{figure}

The $B \to
\eta' X_s$ yields are determined by fitting the observed $\eta'$
signals, after all selection criteria have been applied in the range
$2.0<\petap<2.7\gevc$. We fit separately the $K^{\pm}$ modes and
the \KS modes with results as shown in Fig.~\ref{Fi:semExclsig}. 
We find $188.8\pm21.5$ events for the
$K^{\pm}$ modes and $57.1\pm14.7$ events for the \KS modes in
on-resonance data. For off-resonance data, we find
$0.0^{+2.5}_{-0.0}$ for $K^{\pm}$ modes and $0.0^{+3.0}_{-0.0}$
for \KS modes. 

The efficiencies have been computed with two main
Monte Carlo simulations to study model dependence. One
involves a mixture of resonant modes, $\eta'K$, $\eta' K_1$,
$\eta' K_2^*$, $\eta' K_3^*$, and $\eta' K_4^*$, and the other uses an
$X_s$ pseudo particle decaying to $s \overline{q}g$ $(q=u,d)$.
In the latter, we use a $M(X_s)$ distribution derived from the 
$\eta'$ QCD anomaly theoretical prediction \cite{ref:AtwSon,ref:Ahmady}.

For a given $K$ mode ($K^{\pm}$ or \KS), the efficiency is computed as:
\begin{displaymath}
\epsilon=\frac{N_{\rm fit}}{N_{\rm gen}}
\end{displaymath}
where $N_{\rm fit}$ and $N_{\rm gen}$ are, respectively, the numbers of fit and generated events.
The efficiencies are estimated (statistical errors only) 
to be $(7.6\pm0.4)$\% for $K^\pm$ modes and 
$(3.3\pm0.2)$\% for \KS modes, including the $\KS \to \pi^+\pi^-$ branching fraction. 

\subsection{Semi-inclusive branching fraction and interpretation}
The semi-inclusive rate for the high momentum region,
$2.0<\petap<2.7$\gevc, is computed by performing a weighted average of the results 
obtained in the $K^{\pm}$ modes and the \KS modes.
The detection efficiencies given in Section~\ref{semetaP} are corrected to 
account for the $\eta'$ and $\eta$ branching fractions to the channel we
observe (17.4\%), and the $\Kz\to\KS$ projection (1/2). The final
state $X_s$ includes both \Kp- and \Kz-tagged decays, and both charged and neutral 
$B$ mesons contribute to the observed yields. The total number of produced
$B$ mesons of both charges is 45.5 million.
Computing the weighted average of the \Kp- and \Kz-tagged decays,
assuming that their branching fractions are equal, 
we obtain 
$\mathcal{B}(B \to \eta' X_s)=(6.8^{+0.7}_{-1.0}(stat)\pm1.0(syst)^{+0.0}_{-0.5}(bkg))\times 10^{-4}$.
Our measurement confirms, with higher precision, the CLEO result, 
$(6.2\pm1.6(stat)\pm1.3(syst)^{+0.0}_{-1.5}(bkg))\times 10^{-4}$.

Sources of systematic error included in this result are summarized in
Table~\ref{Ta:semexclsystem}.
The largest uncertainty arises from our ability to model the $X_s$ system. 
This contribution is estimated by determining the efficiency with different Monte Carlo generators,
as described in Section~\ref{semetaP}. 
The efficiency from the $X_s$ pseudo-particle model tends to be lower than that
obtained with the mixture of resonant decays, the difference is smaller when we 
increase the amount of heavier resonances ($K_3^*$, $K_4^*$) in the model.
Other systematic uncertainties are contributed by the 
event shape requirement, $B\overline{B}$ counting, tracking and photon detection efficiencies, 
and kaon identification efficiencies.

\begin{table}[htb]
\caption{Systematic errors for $\mathcal{B}(B \to \eta' X_s)$}
\begin{center}
\begin{tabular}{|l|c|c|}
\hline 
Source  & $K^{\pm}$ modes & \KS modes\\ \hline\hline
Model dependency  & $\pm 10\%$  & $\pm 10\%$  \\
Event shape       & $\pm 1\%$   & $\pm 1\%$   \\
$B$ counting      & $\pm 1.6\%$ & $\pm 1.6\%$ \\
Tracking          & $\pm 5\%$   & $\pm 7\%$   \\
Photon detection  & $\pm 6.8\%$ & $\pm 7.2\%$ \\
$\mathcal{B}(\eta'\to\eta\pi\pi)\times\mathcal{B}(\eta\to\gamma\gamma)$ 
                                & $\pm 3.4\%$ & $\pm 3.4\%$\\
$K^{\pm}$ identification  
                                & $\pm 2.6\%$ &  --           \\
\hline
\end{tabular}
\end{center}
\label{Ta:semexclsystem}
\end{table}

We have also considered a possible contribution from internal spectator
decays $\overline{B}^0 \to \eta'D^{0(*)}$
as an additional source of systematic error. Early branching fraction predictions  for these modes were
in the range $(1.5-6.0)\times 10^{-5}$ \cite{ref:etapd0} while more recent predictions \cite{ref:newetapd0}
find $(3-5)\times 10^{-5}$, assuming that only the quark content of the $\eta'$ contributes to the tree diagram.
The efficiency computed for these modes is 2\%,
which implies a contribution of (1.4-4.6)\% to the observed yield. 
To conservatively account for the theoretical uncertainty,
we use a branching fraction of $9.0\times 10^{-5}$ in determining the systematic error contribution.

Several models have been proposed to explain $\eta'$ production at
high $\petap$ in $B$ decays:
\begin{itemize}
\item{Introduction of $b \to sq\overline{q}$ operators with constructive 
interference between $u\overline{u}$, $d\overline{d}$ and 
$s\overline{s}$ components of the $\eta'$ \cite{ref:Datta}; however 
the expected branching fraction from such processes 
is much lower than we observe.
}
\item{A $b \to c\overline cs$ enhancement through a possible $c\overline{c}$ content 
in the $\eta'$ wave function \cite{ref:Hou,ref:Yuan}; however, more
recent calculations \cite{ref:charmcont} do not support such a wave function
component.
}
\end{itemize}
Both of these mechanisms predict an $X_s$ mass spectrum peaking near 1.5\gevcc,
lower than suggested by the data. The 
branching fraction and mass spectrum in data do appear to be consistent
with a model incorporating an
$\eta'$ QCD anomaly \cite{ref:AtwSon,ref:Ahmady} that couples the 
$\eta'$ to gluons, $b \to sg^*$, $g^* \to g \eta'$. 

\section{Summary}
\label{sec:Summary}
We have measured a semi-inclusive $\eta'$ branching fraction using 
fully reconstructed final states consisting of an $\eta'$ and
a system comprising one kaon and up to four pions
to reduce the background from other $B$ decays. We find
$\mathcal{B}(B \to \eta' X_s)=(6.8^{+0.7}_{-1.7}(stat)\pm1.0(syst)^{+0.0}_{-0.5})\times10^{-4}$
for $2.0<\petap<2.7$\gevc.
This measurement confirms the surprisingly large branching fraction found by 
CLEO, but with much improved precision. This and
the shape of the $X_s$ mass distribution are consistent with the $\eta'$ QCD
anomaly model.

\section{Acknowledgments}
\label{sec:Acknowledgments}

We are grateful for the 
extraordinary contributions of our \pep2\ colleagues in
achieving the excellent luminosity and machine conditions
that have made this work possible.
The collaborating institutions wish to thank 
SLAC for its support and the kind hospitality extended to them. 
This work is supported by the
US Department of Energy
and National Science Foundation, the
Natural Sciences and Engineering Research Council (Canada),
Institute of High Energy Physics (China), the
Commissariat \`a l'Energie Atomique and
Institut National de Physique Nucl\'eaire et de Physique des Particules
(France), the
Bundesministerium f\"ur Bildung und Forschung
(Germany), the
Istituto Nazionale di Fisica Nucleare (Italy),
the Research Council of Norway, the
Ministry of Science and Technology of the Russian Federation, and the
Particle Physics and Astronomy Research Council (United Kingdom). 
Individuals have received support from the Swiss 
National Science Foundation, the A. P. Sloan Foundation, 
the Research Corporation,
and the Alexander von Humboldt Foundation.


\begin{thebibliography}{99}

\bibitem{ref:Neubert}
M. Neubert {\em et al.}, Phys. Rev. Lett. {\bf 81} 5076 (1998), Phys. Lett. B {\bf 41}, 403 (1998).
\bibitem{ref:Ali}
A. Ali {\em et al.}, Phys. Rev. {\bf D59} (1999) 014005.
\bibitem{ref:Cleopap}
CLEO Collaboration, T.E. Browder {\em et al.}, Phys. Rev. Lett. {\bf 81} 1786-1790 (1998).
\bibitem{ref:Datta}
A. Datta, X-G. He, and S. Pakvasa, Phys. Lett. {\bf B 419} 369-376
(1998).
\bibitem{ref:babar}
The \babar\ Collaboration, B. Aubert {\em et al.},
SLAC-PUB-8569, hep-ex/0105044, 
to appear in Nucl.\ Instrum.\ and Methods.
\bibitem{ref:pep2}
\pep2\ Conceptual Design report, SLAC-R-418 (1993).
\bibitem{ref:FoxWolf}
C.G. Fox, S. Wolfram, Nucl. Phys. {\bf B149}, 413 (1979).
\bibitem{ref:AtwSon}
D. Atwood and A. Soni, Phys. Lett. {\bf B 405} 150-156 (1997).
\bibitem{ref:Ahmady}
M.R. Ahmady, E. Kou and A. Sugamoto, Phys. Rev. {\bf D58}, (1998)
014015.
\bibitem{ref:oldetapk}
The \babar\ Collaboration, B. Aubert {\em et al.}, SLAC-PUB-8956, hep-ex/0108017,
submitted to Phys. Rev. Lett. (August 2001).
\bibitem{ref:etapd0}
M. Neubert {\em et al.}, {\it Heavy Flavors}, edited by A.J.Buras and M.Linder, Singapore, 1992; A. Deandrea {\em et al.}, Phys. Lett. {\bf B 318}, 549 (1993).
\bibitem{ref:newetapd0}
A. Deandrea and A.D. Polosa, hep-ph/0107234 (July 2001).
\bibitem{ref:Hou}
W.S. Hou and B. Tseng, Phys. Rev. Lett. {\bf 80} 434-437 (1998).
\bibitem{ref:Yuan}
F. Yuan and K.T. Chao, Phys. Rev. {\bf D56} 2495-2498 (1997).
\bibitem{ref:charmcont}
M. Franz {\em et al.}, Phys. Rev. {\bf D62} (2000) 074024.
\end{thebibliography}
\end{document}